\documentclass[11pt,a4paper]{article}
\usepackage{authblk}
\usepackage{CJK}
\usepackage[utf8]{inputenc}
\usepackage{fullpage}
\usepackage{amsmath,cuted}
\usepackage{amsfonts}
\usepackage[numbers,sort&compress]{natbib}
\usepackage{amssymb}
\usepackage{caption}
\usepackage{floatrow}
\usepackage{chemformula}
\usepackage{graphics}
\usepackage{array}
\usepackage{longtable}
\usepackage{blindtext}

\begin{document}

\title{First principle studies on the optoelectronic properties of rubidium lead halides }
\author[1]{Anupriya Nyayban\footnote{anupriya@rs.phy.student.nits.ac.in}}
\author[1]{Subhasis Panda\footnote{subhasis@phy.nits.ac.in}}
\author[1]{Avijit Chowdhury\footnote{avijit@phy.nits.ac.in}}
\author[2]{B. Indrajit Sharma\footnote{indraofficial@reiffmail.com}}
\affil[1]{Department of Physics, National Institute of Technology Silchar, Assam-788010, India}
\affil[2]{Department of Physics, Assam University, Silchar-788011, India}
\date{}
\maketitle

\begin{abstract}
Entirely inorganic perovskites have attracted enormous attention of late owing to their outstanding applications in optoelectronics including highly stable perovskite solar cells. In-depth understanding of the optoelectronic and transport properties of such materials are vital for practical implementation of the same. The carrier transport properties of the electronic devices based on perovskite materials significantly depend on the effective mass of the respective charge carriers. Here, we have performed first principle calculations with FP-LAPW method for the orthorhombic rubidium lead halide structures (\ch{RbPbX_3}, where \ch{X=I,Br,Cl}) to study the optoelectronic and transport properties. The effective mass of electron (hole) is found to be minimum for \ch{RbPbBr_3} (\ch{RbPbI_3}), suggesting an efficient transport of electrons (holes) in the corresponding materials. Our calculated values such as the dielectric constants, refractive indices, absorption coefficients and reflectivities show good agreement with reported experimental data. To the best of our knowledge, ab-initio study of electronic and optical properties of \ch{RbPbBr_3} \& \ch{RbPbCl_3} in orthorhombic phase (\ch{NH_4CdCl_3} type structure) is reported for the first time.

\end{abstract}
\section{Introduction}
The organic-inorganic halide perovskites have achieved tremendous attention as solar cell materials for the last decade \cite{an29,an30,an31,an32,an33}. The reasons being their strong absorption to the visible range and ability to transport photo-excited electrons and holes with diffusion length longer than the thickness of the film \cite{an9}. The lifetime and effective masses of carries are also high and low respectively for hybrid perovskite. These are required for the material to become an efficient solar cell. The main problem with these hybrid halide perovskites is that they are not thermodynamically stable at ambient conditions. For example, organic-inorganic lead iodide perovskites change to \ch{PbI_2} at atmospheric conditions within hours or days \cite{an10,an11}. Moreover, the dipole due to the organic cation makes the perovskite structure distorted for the hybrid halide perovskites \cite{an12}. These perovskite solar cells (PSC) show current-voltage hysteresis \cite{an13}, which can reduce the stability in PSC and also affects the power conversion efficiency \cite{an34}. These are hindrances to commercialize hybrid halide perovskite solar cells. As the inorganic cation is anisotropic in geometry and less volatile, completely inorganic perovskite solar cell can diminish the hysteresis and hence increases the mobility of electrons \cite{an14}. The replacement of \ch{A} cation in \ch{ABX_3} type structure is a point of interest due to the fact that the stability comes from the A cation. Therefore many efforts are going on to understand the optoelectronic properties of inorganic perovskites due to its high stability. \ch{CsPbX_3}, where \ch{X=I,Br,Cl}, were found to have comparable solar cell device performance as hybrid perovskite \ch{MAPbI_3} while having better stability \cite{an15,an16,an17}. Halide replacement also influences the absorbance as well as the carrier transportation. The variation of different halide changes the mobility from 3 to 5 $\rm{cm^{2}/{Vs}}$ when carrier concentration varies from $10^{15}$ to $10^{18}$ $cm^{-3}$\cite{an18}. \ch{CsSnI_3} is also a promising candidate for the optoelectronic devices having a direct band gap of $1.3 $ eV and mobility of 400 $\rm{cm^2/Vs}$ \cite{an19}.\\The ionic radii of \ch{Cs^+}, \ch{Rb^+} are 1.76 and 1.52 {\AA} producing the Goldschmidt's tolerance factors ($t$) in case of \ch{APbI_3} to be 0.81 and 0.78 respectively \cite{an20}.  These tolerance values are in the range ($0.7 < t < 1$) of stable halide perovskites. Few studies have been reported on \ch{RbPbX_3} structures. A study has been reported on \ch{Rb_{1-x}Cs_xPbI_3} that while increasing the x content of \ch{Cs}, there is a decrease in lattice constants, binding energies, band gaps and carriers' effective masses \cite{an21}. The static dielectric constant has been increased when \ch{Rb} content increases over \ch{CsPbI_3} and indicates the increasing solar cell performance.
The performance of \ch{RbPbI_3} as solar cell was reported in another study with \ch{FTO/TiO_2/RbPbI_3/Spiro-MeOTAD/Au} configuration \cite{an14}. The device was observed to have open circuit voltage as 0.62 V, photo-current density as 3.75 $\rm{mA/cm^2}$, fill factor as 44.60\% and power conversion efficiency as 1.04\% in the direction of reverse sweeping. It reveals the lower performance of \ch{RbPbI_3} as solar cell. Therefore it makes an opportunity to study thoroughly about the optoelectronic properties of \ch{RbPbX_3} so that further investigation can be made to find a highly stable perovskite making it suitable for commercial applications.\\\ch{RbPbI_3} gets the orthorhombic structure of \ch{NH_4CdCl_3} in the $Pnma$ space group and does not change to any other phase before it melts \cite{an8}. At room temperature, \ch{RbPbBr_3} has a mixed phase of \ch{RbPbBr_3} of $Pnma$ and \ch{RbPb_2Br_5} of $I4/mcm$ space group. The room temperature dominant phase (96.7\%) of \ch{RbPbBr_3} has the orthorhombic structure of \ch{NH_4CdCl_3} type. It undergoes transitions to different phases when it is heated till 350$^{\circ}$C. The tetragonal phase of $I4/mcm$ is dominated at 50$^{\circ}$C while at 250$^{\circ}$C both the room temperature one and a distorted perovskite phase of $Pnma$ space group exist \cite{an7}. Like \ch{RbPbBr_3} there is a mixture of \ch{Rb_6Pb_5Cl_16} of $P4/mbm$ and \ch{RbPb_2Cl_5} of $P2_1/c$ space group at room temperature. But \ch{RbPbCl_3} has been found to form a tetragonal phase at 320$^{\circ}$C and cubic phase at 350$^{\circ}$C with $P4/mbm$ and $Pm\bar{3}m$ space groups respectively \cite{an7}.\\In this study we have reported a comparative study of the optoelectronic properties of \ch{RbPbX_3} with the help of first principle calculations. To make a suitable comparison, the stable \ch{NH_4CdCl_3} structure of room temperature has been considered for all. The \ch{NH_4CdCl_3} type structure for \ch{RbPbCl_3} is the hypothetical one. \\The computational method is described in Sec.2. Sec. 3 deals with mainly results obtained in this study. The electronic structures of the studied materials are given in  sub-section 3.1. The detailed description of optical properties and transport properties are presented in sub-section 3.2 and 3.3 respectively. Finally we summarize our observations and conclude in Sec. 4.  
\section{Computational method}
\begin{figure}
\includegraphics[width=9cm,height=6cm]{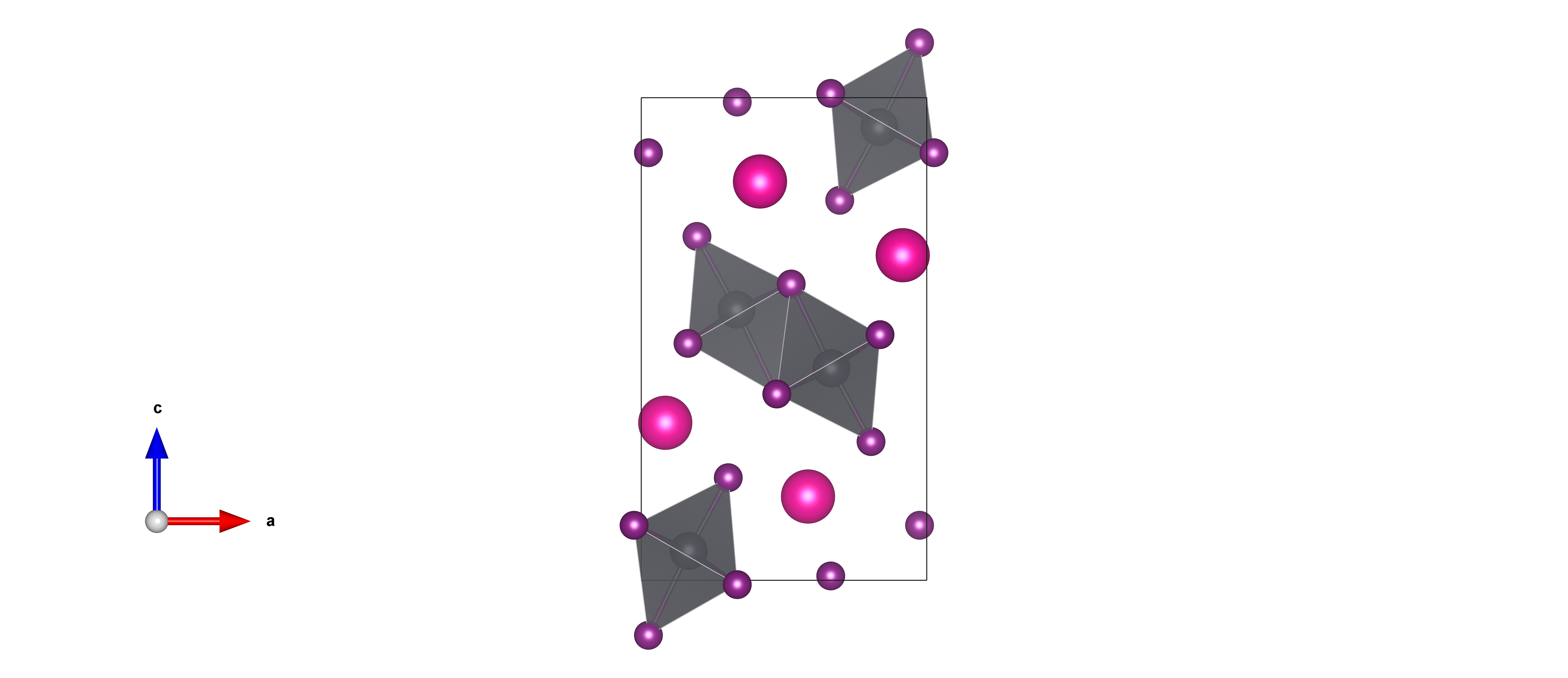}
\includegraphics[width=6cm,height=7cm]{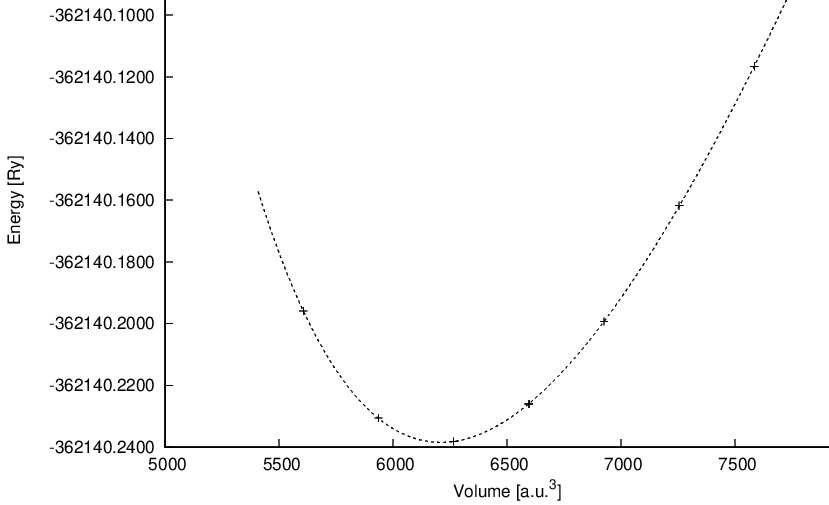}
\centering
\caption{Left: Unit cell of \ch{RbPbX_3}. Magenta, grey and purple spheres represent \ch{Rb}, \ch{Pb} and \ch{X} atoms respectively; Right: Volume optimization for \ch{RbPbI_3}.}
\end{figure}
The present calculations were performed using Wien2k \cite{an22} with the FP-LAPW method. Spin orbit coupling (SOC) has been excluded in the entire calculation. Although Wien2k gives the option of including local orbits in the basis and hence improves linearization and proper orthogonality, we have not included the local orbit in the basis set to reduce the computational cost. There is no shape approximation made for potential and the electronic charge density. GGA based exchange correlation potential is treated with the scheme of Perdew, Burke and Ernzerhof \cite{an23}. Structural unit for all the orthorhombic phases consists of 4 formula units. Values of $ \rm{RK_{max}} $ are set to 9, 8, 8 for \ch{RbPbCl_3}, \ch{RbPbBr_3} and \ch{RbPbI_3} respectively, it determines the size of the basis sets in the calculation. Muffin tin radii have been selected in such a way, that it should not be large enough to leak charges from the sphere and small enough to increase the calculation time. The muffin tin radii are set to 2.5 a.u for all atoms in \ch{RbPbI_3}, 2.31 a.u for all in \ch{RbPbBr_3}; 2.5 a.u. are for \ch{Rb} and \ch{Pb} and 2.2 a.u. for \ch{Cl} in the case of \ch{RbPbCl_3}. Brillouin zone integrations are performed by the tetrahedron method, $6\times13\times3$, $6\times12\times3$ and $6\times13\times3$ k-mesh have been used for \ch{RbPbI_3}, \ch{RbPbBr_3} and \ch{RbPbCl_3} respectively. But a denser k points of $13\times27\times7$ i.e. 392 in the irreducible BZ, are used to estimate the imaginary part of the dielectric tensor during the optical property estimation. Optical properties are calculated with OPTIC \cite{an27} as implemented in Wien2k. The energy and charge convergence criteria  have been fixed  as $0.0001$ Ryd and $0.001e$ respectively. Structural lattice parameters are estimated by fitting energy and volume to the Murnaghan equation of state \cite{an24} for each compound.

\section{Results and discussion}
\subsection{Electronic structure results}
The estimated lattice parameters which are listed in Table 1, show good agreement with the experiments. From Table 1, it is observed that lattice constants decrease from \ch{RbPbI_3} to \ch{RbPbCl_3} as expected \cite{an25}. This is due to the decrease of atomic size of halides from \ch{I} to \ch{Cl}. The optimized volume of \ch{RbPbI_3} can be found from Fig 1. Optimized lattice parameters are calculated from these optimized volumes. The volume optimization graph for \ch{RbPbI_3} is shown only as it is similar for two other materials. 
\\Density of states (DOS) and bandstructure will help to describe the electronic features of \ch{RbPbX_3} compounds. It is observed from the bandstructure that the valence band maximum (VBM) and the conduction band minimum (CBM) occurs at $\Gamma$ and at X respectively, which implies that it is an indirect band gap semiconductor and is well agreed with the similar calculation \cite{an14}. As the band-diagrams are similar for all three structures, in the Fig 2 only the bandstructure of \ch{RbPbBr_3} has been shown. Partial density of states (PDOS) is given in the Figure 2. 6s, 6p orbits of \ch{Pb} and p orbit of halide \ch{X} make the contribution to the VBM while the major contribution comes from p orbit of \ch{X} atoms. 6p of \ch{Pb} and p orbit of halide atoms are responsible for the CBM but it is dominated by p orbit of \ch{Pb} atoms. The lowest lying energy broad peak is centered at $-7.5$ eV and comes mainly from 6s orbital of \ch{Pb} atom. The second lowest lying energy states which have range from 0 to $-2.64$ eV for \ch{RbPbI_3}; 0 to $-2.78$ eV for \ch{RbPbBr_3} and 0 to $-2.83$ eV for \ch{RbPbCl_3} and these are contributed by p orbits of halide \ch{X}, 6p of \ch{Pb} and 6s of \ch{Pb} but it is dominant by p orbits of halide atoms. Above the Fermi level, first broad peak ranges from 2.44 eV to 4.12 eV for \ch{RbPbI_3}; 2.81 eV to 4.77 eV for \ch{RbPbBr_3} and 3.39 eV 5.29 eV for \ch{RbPbCl_3} structure. These states are occupied by the 6p orbit of \ch{Pb} and p of \ch{X} atoms. All peaks are shifted towards the higher energy as we move from \ch{RbPbI_3} to \ch{RbPbCl_3}.\\The calculated band gap values are $2.45$, $2.77$ and $3.27$eV corresponding to \ch{RbPbI_3}, \ch{RbPbBr_3} and \ch{RbPbCl_3} respectively. The band gap increases from \ch{I} to \ch{Cl} in \ch{RbPbX_3} due to the decrease of the halide atomic radii. For the band gap calculation GGA underestimates in comparison with the experimental values. This happens because of the fact that the exchange-correlation term can not be handled accurately \cite{an28}. To the best of our knowledge the electronic structure of \ch{RbPbBr_3} of \ch{NH_4CdCl_3} type structure is calculated for the first time. 
\begin{table}
\caption{Lattice parameters are in Angstrom for \ch{RbPbX_3}, references for experimental parameters for\ch{RbPbI_3} and \ch{RbPbBr_3} are \cite{an8} and \cite{an7} respectively.}
\begin{center}
\begin{tabular}{ccccccc}
\hline
& \multicolumn{3}{c}{Present calculation} & \multicolumn{3}{c}{Experiment}\\
\hline
& a & b & c & a & b & c\\
\hline
\ch{RbPbI_3} & $10.540$ & $4.897$ & $17.817$ & $10.42$ & $4.841$ & $17.615$\\
\ch{RbPbBr_3} & $9.529$ & $4.678$ & $16.678$ & $9.354$ & $4.591$ & $16.372$ \\
\ch{RbPbCl_3} & $9.108$ & $4.471$ & $15.942$ & $-$ & $-$ & $-$ \\
\hline
\end{tabular}
\end{center}
\end{table}

\begin{table}
\caption{The band gap values for \ch{RbPbX_3} are in eV}
\begin{center}
\begin{tabular}{cccc}
\hline
 & Our calculation & Other calculation & experiment\\
\hline
\ch{RbPbI_3} & 2.45 & 2.663\cite{an26} & 2.64\cite{an14}\\
\ch{RbPbBr_3} & 2.77 & $-$ & $-$\\
\ch{RbPbCl_3} & 3.27 & $-$ & $-$\\
\hline
\end{tabular}
\end{center}
\end{table}
\begin{figure}
\includegraphics[width=7cm,height=12cm]{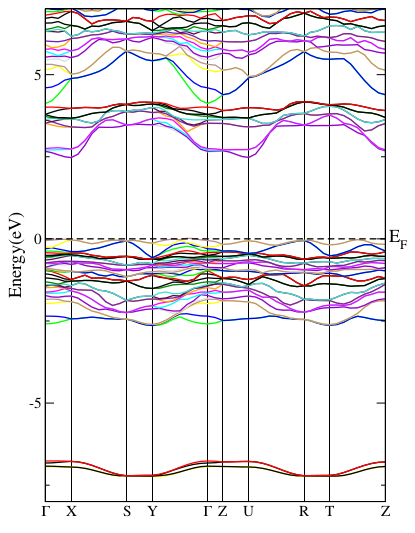}
\includegraphics[width=7cm,height=12cm]{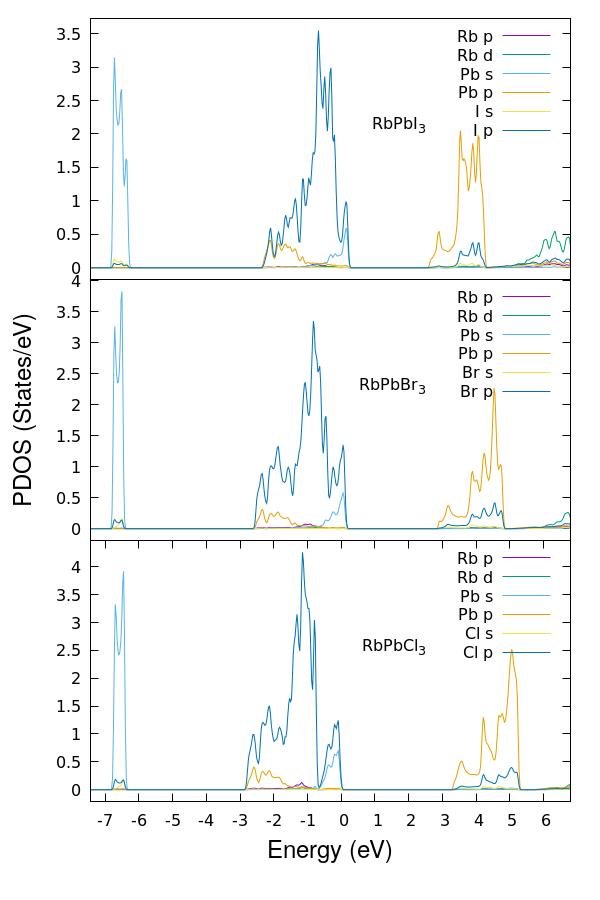}
    \centering
	\caption{Left: bandstructure of \ch{RbPbBr_3}; Right: PDOS for \ch{RbPbI_3}, \ch{RbPbBr_3} and \ch{RbPbCl_3}, Fermi level is set to 0 eV energy. }
\end{figure}


\subsection{Optical properties}
The optical properties of the halide perovskite materials have been studied theoretically from the complex dielectric function $\varepsilon(q,\omega)$, where $q$ is the momentum due to the electron-phonon interaction and $\omega$ is then energy. The momentum transfer is assumed to be zero at lower energy and electric dipole approximation is considered for the optical analysis ~\cite{an1}. The complex dielectric function at lower energy is expressed by 
\begin{equation}
\label{eq1}{\varepsilon(\omega)=\varepsilon_{1}(\omega)+i\varepsilon_{2}(\omega)},
\end{equation} where $\varepsilon_{1}(\omega)$ and $\varepsilon_{2}(\omega)$ are the real and imaginary part of the dielectric function, respectively. There are two types of transitions i.e. intraband and interband, each can be further separated into direct and indirect transitions category. \ch{RbPbX_3} being semiconductor, the intraband contributions is neglected in our calculation. The indirect interband transition is also neglected as it involves with the phonon scattering and its contribution is small \cite{an2}. For calculating the dielectric function, local field effect and finite lifetime effects have been ignored. The direct interband contribution to $\varepsilon_{2}(\omega)$ has been estimated by summing up all the transitions taken place between the occupied and unoccupied states. According to the dipole selection rules, momentum matrix elements allow or forbid interband transitions. Matrix elements are 
\begin{equation}
\label{eq2}{\rm{{M_{G}^{ln}}(k,q)=\langle \psi_{k-ql} \vert e^{-i(q+G)r} \vert \psi_{kn} \rangle}}
\end{equation}
A detailed description about the matrix is given in the following references \cite{an3,an4}. The off-diagonal terms of the dielectric tensor $\varepsilon_{2}$ are negligible compared to the diagonal elements due to the fact that principal axes make an orthogonal system. The diagonal elements of the tensor $\varepsilon_{2}$ is 
\begin{equation}
\label{eq3}{\varepsilon_{2}^{ss}={\frac{{8 \pi^{2} e^{2}}}{{m^{2} \omega^{2}}}}\sum_{n}^{unocc} \sum_{n'}^{occ} \int_{BZ} |P_{nn'}^{s}(k)|^{2} f_{kn}(1-f_{kn'})\times \delta(E_{n}^{k} - E_{n'}^{k} - \hbar \omega) \times {\frac{{d^{3}k}}{{(2\pi)^{3}}}}},
\end{equation}
where $m$ and $e$ are mass and charge of electrons, $f_{kn}$ is the Fermi-Dirac distribution function, $P_{nn'}^{s}$ is the projection of the momentum matrix along $s$ i.e. $x,y,z$ directions and $E_{n}^{k}$ is the single electron energy. Matrix elements are calculated over the muffin-tin as well as the interstitial regions. The total $\varepsilon_{2}^{ss}$ has been estimated over the irreducible Brillouin zone (IBZ) as 
\begin{equation}
\label{eq4}
{\varepsilon_{2}^{ss}=\frac{{1}}{{N}}\sum_{i=1}^{N} \sigma_{i}^{T} \varepsilon_{2}(IBZ) \sigma_{i}},
\end{equation} 
where $\sigma_{i}$ and $N$ are the symmetry operations and the number of symmetry operations respectively. Using Kramer-Kronig relation, the real part of the dielectric tensor $\varepsilon_{1}$ can be estimated as 
\begin{equation}
\label{eq5}
{\varepsilon_{1}(\omega)=1+\frac{{2}}{{\pi}} \int_{0}^{\infty} {\varepsilon_{2}(\omega')\omega' {d\omega'}}/{(\omega'^{2}-\omega^{2})}}.
\end{equation}
With the help of real and imaginary part of dielectric function, refractive index and extinction coefficient can be evaluated as follows 
\begin{equation}
\label{eq6}
{n(\omega)=\frac{1}{\sqrt{2}} \big[\sqrt{\varepsilon_{1}^{2}(\omega)+\varepsilon_{2}^{2}(\omega)}+\varepsilon_{1}(\omega)\big]^{{1}/{2}}},
\end{equation}
\begin{equation}
\label{eq7}
{k(\omega)=\frac{1}{\sqrt{2}} \big[\sqrt{\varepsilon_{1}^{2}(\omega)+\varepsilon_{2}^{2}(\omega)}-\varepsilon_{1}(\omega)\big]^{{1}/{2}}}.
\end{equation}
Further reflectivity ($R$) due to normal incidence, absorption coefficient ($\alpha$) and ELF ($L$) can be expressed as a function of the extinction coefficient and refractive index using
\begin{equation}
\label{eq8}
{R={\frac{(n-1)^{2}+k^{2}}{(n+1)^{2}+k^{2}}},}
\end{equation}
\begin{equation}
\label{eq9}
{\alpha=\frac{2\omega k}{c}},
\end{equation}
\begin{equation}
\label{eq10}
{L=-Im\left(\frac{1}{\varepsilon_{1}+i\varepsilon_{2}}\right)=\frac{\varepsilon_{2}}{\varepsilon_{1}^{2}+\varepsilon_{2}^{2}}},
\end{equation}
where $\varepsilon_{1}$, $\varepsilon_{2}$, $n$, $k$ are used for $\varepsilon_{1}(\omega)$, $\varepsilon_{2}(\omega)$, $n(\omega)$, $k(\omega)$ respectively and $c$ is the velocity of light in free space.

\begin{figure}
	\includegraphics[width=7cm,height=11cm]{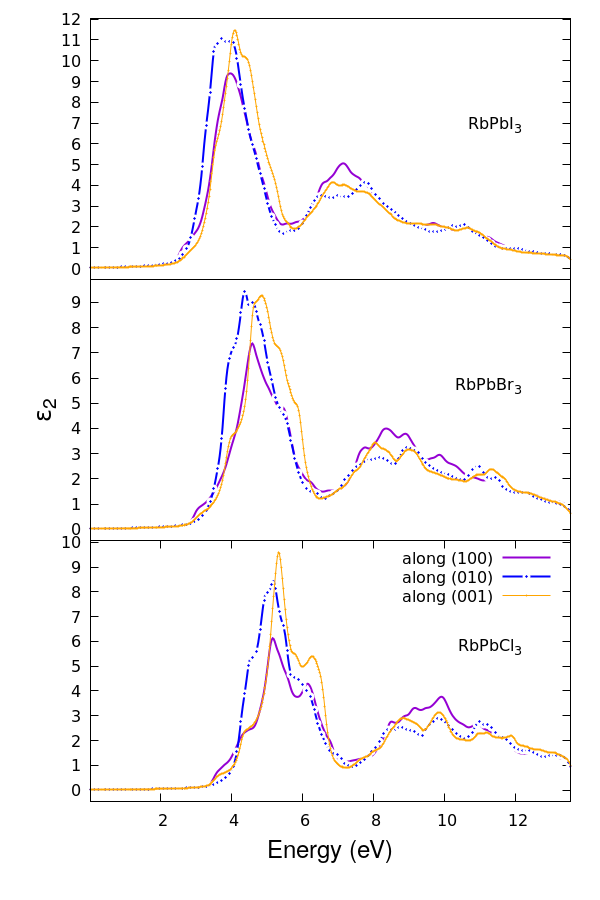}
	\includegraphics[width=7cm,height=11cm]{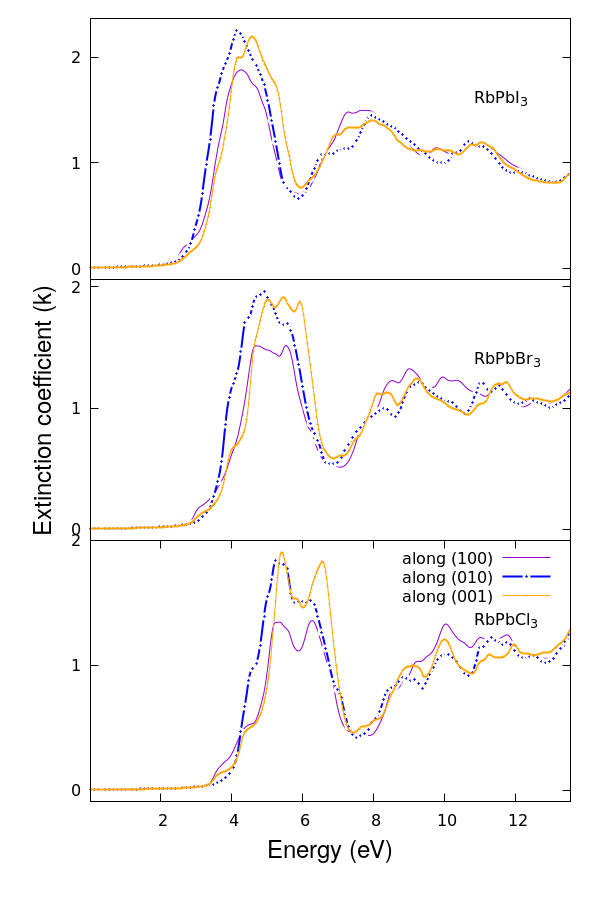}
	\includegraphics[width=7cm,height=11cm]{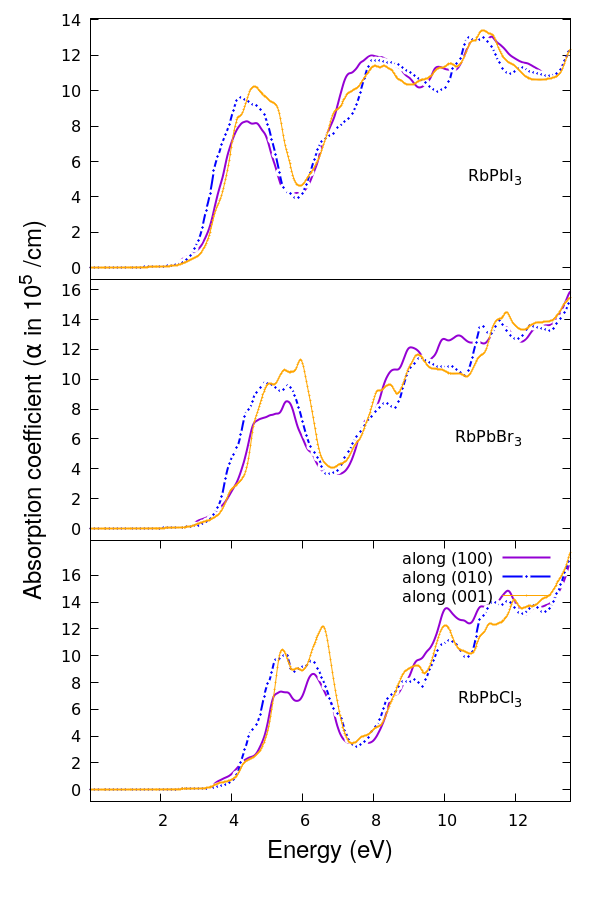}
	\centering
	\caption{Top left: Imaginary part of the dielectric function $\varepsilon_{2}$; Top right: Extinction coefficient k; Bottom: Absorption coefficient $\alpha$ as a function of photon energy for \ch{RbPbX_3} where \ch{X} runs from \ch{I} to \ch{Cl} along the three polarization directions}
\end{figure}
\begin{figure}
\includegraphics[width=7cm,height=11cm]{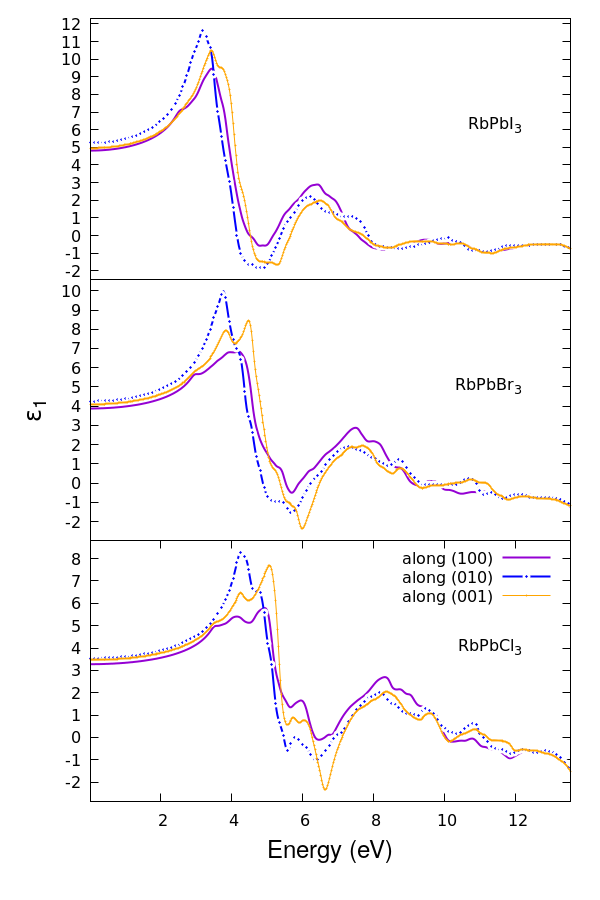}
\includegraphics[width=7cm,height=11cm]{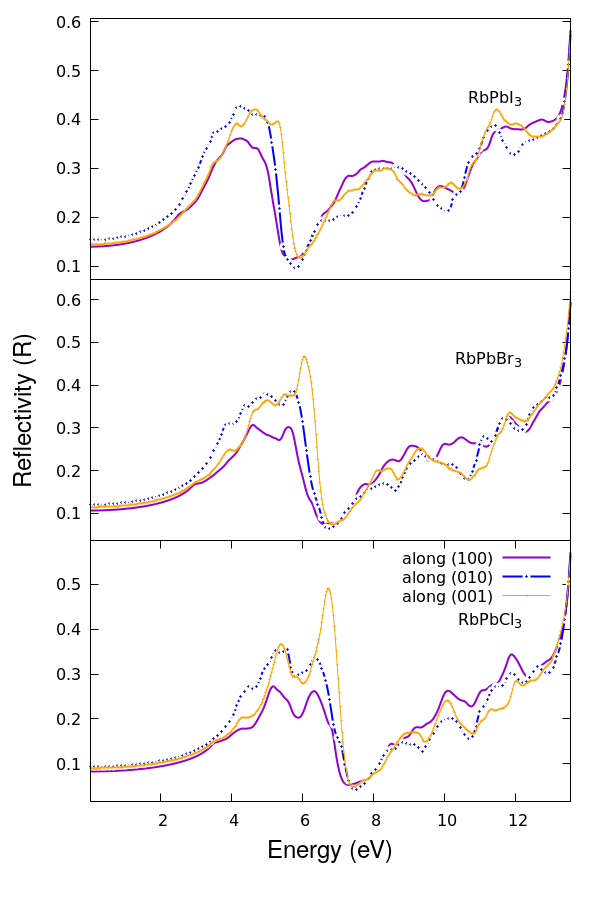}
\includegraphics[width=7cm,height=11cm]{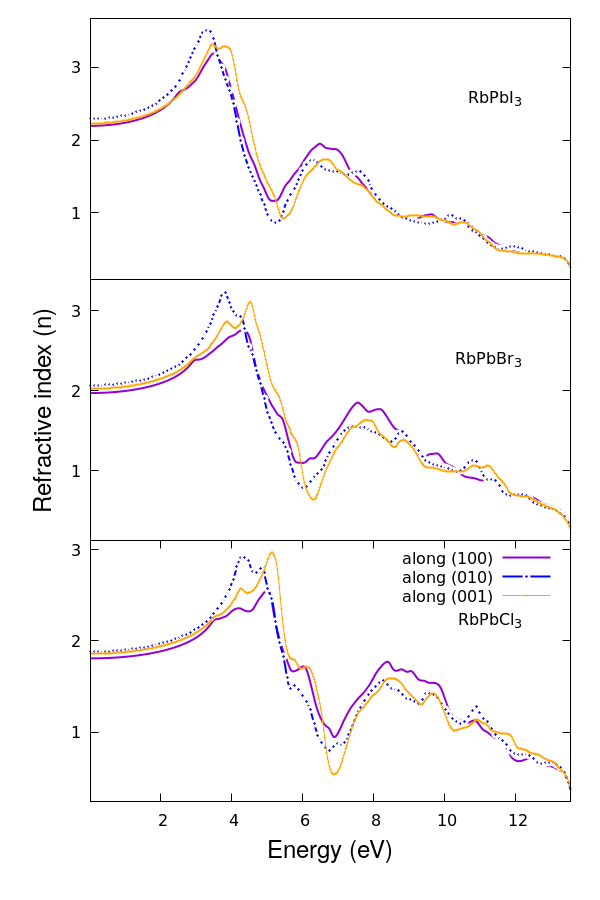}
\includegraphics[width=7cm,height=11cm]{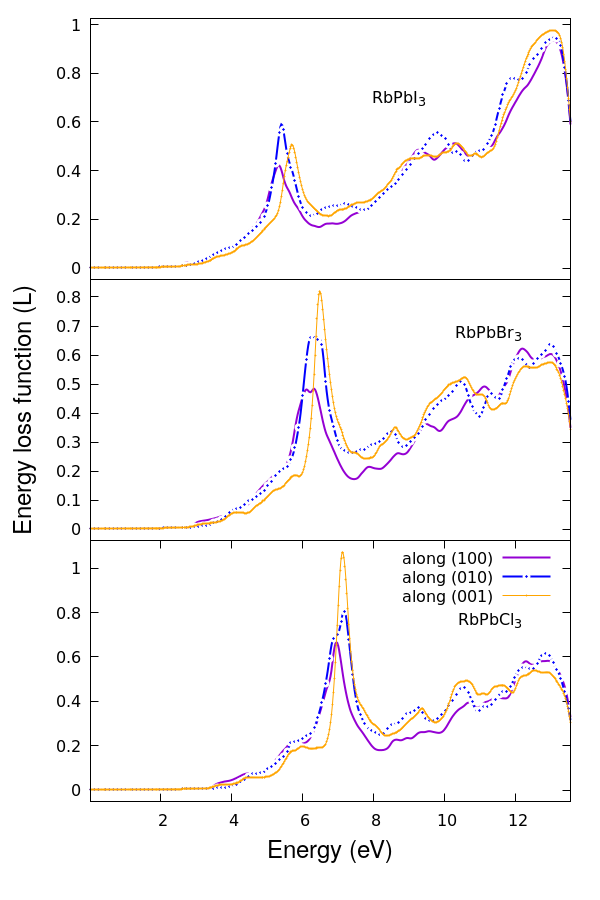}
\centering
\caption{Top left: the real part of the dielectric function $\varepsilon_1$; Top right: Reflectivity R; Bottom left: Refractive index n; Bottom right: Energy loss function L for \ch{RbPbI_3}, \ch{RbPbBr_3}, \ch{RbPbCl_3}}
\end{figure}
The large  number of sub-bands are involved for the orthorhombic system compared to the cubic one. So, the interband transitions as well as peak positions for $\varepsilon_{2}$ spectra corresponding to ${k}$ of DOS are difficult to analyze. Therefore, we have preferred the analysis of the interband transitions corresponding to the main peaks of $\varepsilon_{2}$ for all the structures at the low energy region along the $x$, $y$, $z$ polarization directions. It can be seen from the figure 3 that the nature of $\varepsilon_2$ is similar for all \ch{RbPbX_3} structures. It is observed from the Fig 2 that the VB width is narrower for \ch{RbPbI_3} as compared to \ch{RbPbBr_3} and \ch{RbPbCl_3}. Moreover, \ch{RbPbI_3} displays higher peak intensity and smaller peak width in the $\varepsilon_{2}$ spectra. The spectra are also shifted towards the higher energy region as the halide is changed from \ch{I} to \ch{Cl} in \ch{RbPbX_3}. It is due to the increase of the band gap from \ch{I} to \ch{Cl}. Projected density of states will predict states responsible for transitions corresponding to a particular peak in the spectra of $\varepsilon_2$. Three peaks are observed. The first peak is due to the transition of electrons from a mixed states of \ch{(I,Br,Cl)}-p, \ch{Pb}-6s  to 6p conduction state of \ch{Pb}. The second and third peaks are observed due to the electronic transitions from \ch{(I,Br,Cl)}-p valence states to \ch{Rb}-4d unoccupied conduction states and  \ch{Pb}-6s to a mixed states of \ch{Pb}-6p and \ch{(I,Br,Cl)}-p states respectively. Similar kind of transitions have been observed for \ch{CsPbX_3} \cite{an25}. The spectrum are shifted towards the higher energy region as the halide changes from \ch{I} to \ch{Cl}. Extinction coefficient spectra also displays the similar behavior as $\varepsilon_2$. It is observed from the Fig 3 that absorption coefficients have higher peaks for (001) polarization direction till 7 eV energy of the spectra for all structures. The reason is that  the extinction coefficients along (001) directions have high peaks as along the (010) directions while peak positions are in the higher energy region for (001) direction.\\The spectra of real part of the dielectric function have been shown in the figure 4. At frequency zero, the value of $\varepsilon_1(0)$ is the static dielectric constant. Fig 4 shows that $\varepsilon_1(0)$ decreases from \ch{RbPbI_3}, \ch{RbPbBr_3} and then to \ch{RbPbCl_3}. It proves the inverse relationship with the band gap. Similar feature has been observed for other materials \cite{an5,an6}. The calculated dielectric constants are matching well with the previous report \cite{an26} where values are 5.0, 4.4 and 4.3 along (001), (010) and (001) polarization directions respectively in the case of \ch{RbPbI_3}. The dielectric constant value is closer to that of a compound having an active lone pair (5.64) and these Pb-halide based materials can be polarized to an external field \cite{an26}. $\varepsilon_1$ increases from zero photon energy, it attains maximum value and then decreases. At some frequency, it goes below zero, indicating the metallic behavior. The real part of the static dielectric constants $\varepsilon_1(0)$ decreases from \ch{I} to \ch{Cl}. All values are higher along (010) directions. Values of $\varepsilon_1(0),n(0),R(0)$ along (100), (010), (001) polarization directions are listed in Table 3. The reflectivity starts from 14\%, 15.4\% and 14.4\% for \ch{RbPbI_3} along (100), (010), (001) directions respectively. While for \ch{RbPbBr_3} and \ch{RbPbCl_3} these values are 10.7\%, 12\% 11.5\% and 8.3\%, 9.3\%, 9\% respectively. Starting reflectivity decreases from \ch{I} to \ch{Cl} as reported in \cite{an25}. Sharp peaks in the energy loss function $\rm{L(\omega)}$ spectra provide the plasma frequency. This frequency is the result of oscillations of electrons in valence states of the crystal. Peaks in $\rm{L(\omega)}$ spectra are located at those energies corresponding to which, reflectivity R has minimum amplitudes as shown in the figure 4. The ELF is higher for \ch{RbPbCl_3} near 7 eV compared to that of \ch{RbPbBr_3} and \ch{RbPbI_3}. It happens because of the fact that \ch{RbPbCl_3} has the lowest minimum near 7 eV for  the spectra of $\varepsilon_{2}$.  

\begin{table}
\caption{Static values for $\varepsilon_{1}$, R, n for \ch{RbPbX_3} along different polarization directions.}
\begin{center}
\begin{tabular}{cccccccccc}
\hline
\multicolumn{4}{c}{\ch{RbPbI_3}} & \multicolumn{3}{c}{\ch{RbPbBr_3}} & \multicolumn{3}{c}{\ch{RbPbCl_3}}\\
\hline
Along & $\varepsilon_{1}(0)$ & R(0) & n(0) & $\varepsilon_{1}(0)$ & R(0) & n(0) & $\varepsilon_{1}(0)$ & R(0) & n(0) \\
\hline
(100) & 4.8135 & 0.140 & 2.194 & 3.877 & 0.107 & 1.969 & 3.275 & 0.091 & 1.810\\
(010) & 5.248 & 0.154 & 2.290 & 4.262 & 0.121 & 2.064 & 3.545 & 0.094 & 1.883\\
(001) & 4.953 & 0.144 & 2.226 & 4.093 & 0.115 & 2.023 & 3.464 & 0.083 & 1.861\\
\hline 
\end{tabular}
\end{center}
\end{table}

\subsection{Transport properties}
\begin{table}
\caption{The calculated effective masses of electrons and holes for \ch{RbPbX_3} along the high symmetric directions.}
\begin{tabular}{ccccccc}
\hline
Crystallographic directions & \multicolumn{2}{c}{\ch{RbPbI_3}} & \multicolumn{2}{c}{\ch{RbPbBr_3}} & \multicolumn{2}{c}{\ch{RbPbCl_3}}\\
\hline
 & electron & hole & electron & hole & electron & hole \\
\hline
$\Gamma \longrightarrow X$ & $0.245$ & $0.098$ & $0.128$ & $0.112$ & $0.511$ & $0.095$\\
$\Gamma \longrightarrow S$ & $0.143$ & $0.451$ & $0.123$ & $0.298$ & $0.189$ & $0.386$\\
$X \longrightarrow S$ & $0.026$ & $0.152$ & $0.030$ & $0.180$ & $0.037$ & $0.386$ \\
$\Gamma \longrightarrow Z$ & $0.119$ & $0.030$ & $0.056$ & $0.029$ & $0.056$ & $0.035$ \\
$\Gamma \longrightarrow U$ & $0.065$ & $0.053$ & $0.065$ & $0.054$ & $0.076$ & $0.065$ \\
$Z \longrightarrow U$ & $0.293$ & $0.331$ & $0.446$ & $1.321$ & $2.732$ & $2.849$ \\
\hline
\end{tabular}
\end{table}
In order to find the transport behaviour of photo-generated electrons and holes, effective mass of electrons and holes are calculated along the high symmetry directions. The effective mass has been calculated by using the relation
\begin{equation}
\label{eq11}
{\frac{1}{m_{\rm{eff}}}={\frac{1}{\hbar^{2}}}{\frac{\partial^{2}E}{\partial k^{2}}}}.
\end{equation}
The effective masses for electrons and holes have been estimated by fitting the energy and $k$ values corresponding to the lowest conduction band and the highest of the valence band obtained from the band-diagrams respectively. Estimated effective masses in terms of the free electron mass are represented in Table 4. The average effective masses of electrons and holes for \ch{RbPbI_3} are 0.1485 and 0.185 respectively. Average effective masses of electrons and holes for \ch{RbPbBr_3} and \ch{RbPbCl_3} are 0.141, 0.332; 0.600 and 0.636 respectively. \ch{RbPbI_3} has smallest effective masses for electrons and holes compared to other two structures which suggests highest mobility as well as longest diffusion lengths. For \ch{RbPbI_3} the effective mass of electron is smaller than that of  hole. This proves that electrons can be transported easily than holes. Another study has also been reported that \ch{RbPbI_3} can transfer electrons to \ch{TiO_2} \cite{an26}, because, the CBM of \ch{RbPbI_3} is higher than that of \ch{TiO_2}. Because of the flat nature of the band diagram, the effective masses are higher compared to that of the hybrid lead halide perovskites.

\section{Conclusion}
In summary, we have explored the optoelectronic properties of orthorhombic rubidium lead halide (\ch{RbPbX_3}, where \ch{X=I,Br,Cl}) and compared various properties such as dielectric function. absorption coefficient, refractive index and transport properties with the help of first principle calculations. Electronic band structure calculations imply that these materials are wide band gap semiconductors with lowest band gap of 2.45 eV (for \ch{RbPbI_3}). It is evident from the PDOS plot that the VBM of \ch{RbPbX_3} is dominantly occupied by p orbits of halide atoms while the CBM is populated by p orbits of \ch{Pb} atoms. Calculated average dielectric constants (5.005, 4.077 and 3.428) and refractive indices (2.237, 2.019 and 1.851) decrease as the halide changes from \ch{I} to \ch{Cl}. Moreover, the static values of dielectric constants, reflectivity and refractive indices of all the \ch{RbPbX_3} display anisotropic behavior with maximum along [001] polarization direction. The effective masses of electron and hole are found to be minimum for \ch{RbPbBr_3} and \ch{RbPbI_3}, indicating an efficient transport of electrons/ holes in the corresponding materials, respectively. The result from the first principle calculations indicate that among all three rubidium lead halide structures, \ch{RbPbI_3} is the most promising one for the photovoltaic applications due to lower band gap energies. However, further study on the band gap tuning through suitable doping is required for its potential application in solar cells.
  
\bibliographystyle{unsrt}
\bibliography{reference}

\end{document}